\documentclass{ewass_ss4proc}

\usepackage{kantlipsum}
\usepackage{seqsplit}

  \def \teff {$T_{\mathrm{eff}}$}

\editors{Emeline Bolmont \& Sergi Blanco-Cuaresma}
\editorsurl{www.emelinebolmont.com | www.blancocuaresma.com/s/}
\publisher{Zenodo}
\conference{EWASS Special Session 4 (2017): Star-planet interactions}
\conferencedate{2017}

\title{How alien can alien worlds be?}
\author{V. Adibekyan,$^{1}$
	P. Figueira,$^{1}$ 
	N. C. Santos$^{1,2}$}

\affiliation{$^{1}$ Instituto de Astrof\'isica e Ci\^encias do Espa\c{c}o, Universidade do Porto, CAUP, Rua das Estrelas, 4150-762 Porto, Portugal \\
			$^{2}$ Departamento de F\'isica e Astronomia, Faculdade de Ci\^encias, Universidade do Porto, Rua do Campo Alegre, 4169-007 Porto, Portugal}

\shorttitle{Aliens}
\shortauthors{Adibekyan et al.}

\abs{In an attempt to select stars that can host planets with characteristics similar to our own, we selected seven solar-type stars known to host planets in the habitable zone and 
for which spectroscopic stellar parameters are available. For these stars we estimated 'empirical' abundances of O, C, Mg and Si, which in turn we used to derive the
iron and water mass fraction of the planet building blocks with the use of the model presented in \citet{Santos-15}.
Our results show that if rocky planets orbit these stars they might have significantly different compositions between themselves and different from that of our Earth.
However, for a meaningful comparison between the compositional properties of exoplanets in the habitable zone and our own planet, a far more sophisticated 
analysis \citep[e.g.][]{Dorn-17} of a large number of systems with precise mass and radius of planets, and accurate  chemical abundances of the host stars.
The work presented here is merely the first humble step in this direction.}

\begin{document}

\maketitle

\section{Introduction}

Observations revealed a now well-known dependence between exoplanet formation and metallicity. Giant planets tend to form more frequently around metalic stars \citep[e.g.][]{Gonzalez-97, Santos-01, Mortier-13}.
This dependence, however, is less clear for low-mass/small-size planets \citep[e.g.][]{Sousa-11,Buchhave-15, Zhu-16}. Interestingly, there are no planets observed around very metal poor stars
e.g. [Fe/H] < -1 dex (exoplanet.eu), which probably means that there is a critical metallicity below which no planet can be formed \citep[e.g.][]{Johnson-12}. This critical metallicity 
is much higher than the metallicity of population III stars in our Galaxy, leading to the inference  that planet formation started only after the first stars were formed and died, enriching
the the interstellar gas with metals. However, this process did not take very long (in astronomical timescale) since many planets are found around thick disk stars that
are typically older than 8 Gyr \citep[e.g.][]{Haywood-13}. Moreover, it was shown that planet formation was more efficient around thick disk stars when compared to the thick disk stars of the same (low)metallicity
\citep[][]{Haywood-09, Adibekyan-12a, Adibekyan-12b}. This stems from the thick disk stars being enhanced in $\alpha$-elements such as O, Mg, Si \citep[e.g.][]{Bensby-13, Adibekyan-13b} 
which seems to compensate the lack of iron, typically used as a proxy of overall metallicity \citet[][]{Adibekyan-12a}. Indeed a system of five sub-earth-size planets was detected 
around a 11.2 Gyr old star \citep{Campante-15}, setting the limit for the earliest exoplanet system formed and opening a possibility for the existence of ancient life in our Galaxy.

During (at least) the 11.2 billion year-long history of exoplanet formation in the Milky Way, the interstellar gas has chemically evolved significantly. Some recent works detail 
 how abundances of different chemical elements changes with time \citep[e.g.][]{Nissen-17, DelgadoMena-17} and place in the Galaxy \citep[e.g.][]{Recio-Blanco-14, Kordopatis-15}.
Abundances of these different individual heavy elements and specific elemental ratios (e.g. Mg/Si and Fe/Si) are, in turn, very important for the 
formation \citep{Santos-01, Suarez-andres-17, Santos-17, Adibekyan-15, Adibekyan-17}, orbital architecture \citep{Adibekyan-13, Beauge-13, Mulders-16}, structure and 
composition \citep{Santos-15, Thiabaud-14, Dorn-15}, and even maybe for 'habitability' of the exoplanets \citep{Adibekyan-16}. This discussion leads to a conclusion that the chemical environment i.e., 
time and place in the Milky Way, play a crucial role for the formation of planets and their main characteristics \citep{Adibekyan-17b}. 

In a recent work, \citet{Adibekyan-16} proposed that planets in the habitable zone of solar-like stars may have different compositions from that of our Earth. In this work, we try to estimate
the composition of the planet building blocks around stars that are known to host planets in the habitable zone (HZ)\footnote{Do not mix with the Habeertable Zone defined in \citet{Turbo-King-17}.}.

\begin{table*}[t]
	\begin{center}
	\caption{Stellar parameters and abundances of the sample stars. The index $emp$ refer to the 'empirical' derivation of the abundances. The references for stellar parameters
	are in the last column.}
	\label{tab:table1}
	\footnotesize
	\setlength{\tabcolsep}{0.13em}
	\begin{tabular*}{\linewidth}{l c c c c c c c c c c c l}
	\noalign{\smallskip}\hline\hline\noalign{\smallskip}
	star & \teff & $\log$ g & [Fe/H] & [O/H] & [C/H] & [Mg/H] & [Si/H] & [O/H]$_{emp}$ & [C/H]$_{emp}$ & [Mg/H]$_{emp}$ & [Si/H]$_{emp}$ & References \\ 
	\noalign{\smallskip}\hline\noalign{\smallskip}
	Kepler-22 & 5518$\pm$44 & 4.44$\pm$0.06 & -0.29$\pm$0.06 & -0.19$\pm$0.05 & -0.24$\pm$0.05 & -0.23$\pm$0.03 & -0.24$\pm$0.02 & -0.10$\pm$0.10 & -0.28$\pm$0.06 & -0.18$\pm$0.06 & -0.21$\pm$0.05 & \citet{Borucki-12} \\ 
	HD40307 & 4774$\pm$77 & 4.42$\pm$0.16 & -0.36$\pm$0.02 &  & -0.36$\pm$0.10 & -0.20$\pm$0.09 & -0.19$\pm$0.08 & -0.14$\pm$0.10 & -0.40$\pm$0.05 & -0.22$\pm$0.03 & -0.27$\pm$0.05 & \citet{Tsantaki-13} \\ 
	HD10700 & 5310$\pm$17 & 4.46$\pm$0.03 & -0.52$\pm$0.01 & -0.26$\pm$0.10 & -0.52$\pm$0.10 & -0.30$\pm$0.06 & -0.36$\pm$0.01 & -0.18$\pm$0.10 & -0.52$\pm$0.10 & -0.29$\pm$0.08 & -0.35$\pm$0.07 & \citet{Sousa-08} \\ 
	Kepler-452 & 5757$\pm$85 & 4.32$\pm$0.09 & 0.21$\pm$0.09 &  &  &  &  & 0.21$\pm$0.10 & 0.17$\pm$0.08 & 0.22$\pm$0.08 & 0.21$\pm$0.08 & \citet{Jenkins-15} \\ 
	Kepler-62 & 4925$\pm$70 & 4.68$\pm$0.04 & -0.37$\pm$0.04 &  &  &  &  & -0.14$\pm$0.10 & -0.39$\pm$0.05 & -0.21$\pm$0.05 & -0.22$\pm$0.06 & \citet{Borucki-13} \\ 
	Kepler-174 & 4880$\pm$126 & 4.68$\pm$0.15 & -0.43$\pm$0.10 &  &  &  &  & -0.18$\pm$0.10 & -0.59$\pm$0.10 & -0.41$\pm$0.01 & -0.40$\pm$0.04 & \citet{Rowe-14} \\ 
	Kepler-443 & 4723$\pm$100 & 4.62$\pm$0.10 & -0.01$\pm$0.10 &  &  &  &  & 0.02$\pm$0.10 & -0.11$\pm$0.09 & -0.07$\pm$0.10 & 0.06$\pm$0.07 & \citet{Torres-15} \\
	\noalign{\smallskip}\hline
	\end{tabular*}
	\end{center}
	\vspace{-0.2cm}
\end{table*}

\section{Planets in the habitable zone: sample selection}

To select stars with HZ planets we used the the Habitable Exzoplanet Catalog\footnote{http://phl.upr.edu/projects/habitable-exoplanets-catalog}. From the list of 
``Conservative'' and ``Optimistic Sample of Potentially Habitable Exoplanets'' we selected planets that are hosted by solar-type stars with effective temperature higher than 4500 K. We note that 
the derivation of stellar parameters, including stellar metallicity,  is very challenging for cooler stars and are typically less precise. For six 
(Kepler-1540, Kepler-1544, Kepler-1552, Kepler-1090, Kepler-1606 and Kepler-1638) out of 13 selected systems, the stellar metallicity (the most important parameter for the current study)
was derived by \citet{Morton-16} using the \textit{vespa}\footnote{https://github.com/timothydmorton/vespa} package. The non-spectroscopic metallicities of these six stars seem to be too biased 
(probably because of the fitting priors and algorithm) towards the solar value. The mean metallicity and the standard deviation of the six stars is 0.000$\pm$0.058 dex. In fact, 
about 80\% of the stars from the full sample of \citet{Morton-16} have metallicities from -0.1 to 0.1 dex. For comparison, only about 34\% of stars from the volume-limited HARPS sample of 
\citet{Adibekyan-12c} lie within the aforementioned range of metallicity. Without making any judgment on the quality of this work, but nonetheless noting the clear discrepancy, we preferred
to confine our analysis to spectroscopicly derived parameters. As such, we limited our analysis to seven stars 
(Kepler-22, HD40307, HD10700, Kepler-452, Kepler-62, Kepler-174, Kepler-443) with metallicities only derived by spectroscopic methods (see Table\,\ref{tab:table1}). It is interesting 
to see that six out of seven hosts have sub-solar metallicities, although the metallicity of Kepler-443 is compatible with the solar value within the error.

\begin{table*}[t]
	\begin{center}
	\caption{Mass fractions and  total fraction (Z) of heavy elements, iron mass fraction among refractory species ($f_{iron}$), and the water mass fraction ($w_{f}$). All values are in \%.}
	\label{tab:table2}
	\begin{tabular*}{\textwidth}{lcccccccc}
	\hline\hline
	star & H$_{2}$O & CH$_{4}$ & Fe & MgSiO$_{3}$ & Mg$_{2}$SiO$_{4}$ & Z & $f_{iron}$ & $w_{f}$ \\ 
	\hline
	HD40307$_{emp}$ & 0.39$\pm$0.11 & 0.15$\pm$0.01 & 0.06$\pm$0.00 & 0.08$\pm$0.03 & 0.07$\pm$0.02 & 0.74$\pm$0.12 & 27.89$\pm$1.72 & 65.00$\pm$11.58  \\
	HD40307 & 0.38$\pm$0.11 & 0.15$\pm$0.03 & 0.06$\pm$0.00 & 0.10$\pm$0.06 & 0.07$\pm$0.07 & 0.76$\pm$0.12 & 25.10$\pm$2.82 &  62.30$\pm$14.35 \\
	HD10700$_{emp}$ & 0.37$\pm$0.11 & 0.11$\pm$0.03 & 0.04$\pm$0.00 & 0.06$\pm$0.05 & 0.07$\pm$0.05 & 0.65$\pm$0.11 & 23.80$\pm$2.51 & 68.52$\pm$13.08 \\
	HD10700 & 0.38$\pm$0.10 & 0.12$\pm$0.03 & 0.04$\pm$0.00 & 0.06$\pm$0.02 & 0.06$\pm$0.03 & 0.66$\pm$0.10 & 24.36$\pm$1.36 & 70.37$\pm$10.63 \\
	Kepler-22$_{emp}$ & 0.45$\pm$0.13 & 0.20$\pm$0.02 & 0.07$\pm$0.01 & 0.10$\pm$0.05 & 0.08$\pm$0.05 & 0.90$\pm$0.14 & 29.85$\pm$3.24 & 64.29$\pm$14.83 \\
	Kepler-22 & 0.33$\pm$0.04 & 0.20$\pm$0.02 & 0.08$\pm$0.01 & 0.10$\pm$0.02 & 0.07$\pm$0.02 & 0.78$\pm$0.05 & 30.64$\pm$2.90 & 56.90$\pm$5.00 \\
	Kepler-62$_{emp}$ & 0.40$\pm$0.10 & 0.15$\pm$0.02 & 0.06$\pm$0.01 & 0.11$\pm$0.04 & 0.06$\pm$0.03 & 0.76$\pm$0.10 & 25.88$\pm$2.53 & 63.49$\pm$11.22 \\
	Kepler-174$_{emp}$ & 0.39$\pm$0.11 & 0.10$\pm$0.02 & 0.04$\pm$0.01 & 0.08$\pm$0.01 & 0.03$\pm$0.01 & 0.63$\pm$0.11 & 26.86$\pm$4.68 & 72.22$\pm$11.14 \\
	Kepler-443$_{emp}$ & 0.56$\pm$0.14 & 0.29$\pm$0.05 & 0.13$\pm$0.03 & 0.23$\pm$0.04 & 0.02$\pm$0.05 & 1.25$\pm$0.16 & 32.70$\pm$6.00 & 59.57$\pm$15.68 \\
	Kepler-452$_{emp}$ & 0.82$\pm$0.21 & 0.55$\pm$0.10 & 0.21$\pm$0.05 & 0.28$\pm$0.15 & 0.15$\pm$0.15 & 2.02$\pm$0.25 & 32.73$\pm$5.84 & 56.16$\pm$30.27 \\
	\noalign{\smallskip}\hline
	\end{tabular*}
	\end{center}
	\vspace{-0.2cm}
\end{table*}

\section{Abundances of the host stars}

In order to derive composition of the planetary building blocks, as it was done in \citet{Santos-15} chemical abundances of O, C, Mg, and Si are necessary.
Our intensive literature search for chemical abundances of the sample stars was not very productive. Only three stars have elemental abundances reported in the literature: 
Kepler-22 -- \citep{Schuler-15}, HD40307 -- \citep{DelgadoMena-17a, Suarez-andres-17}, and HD10700 -- \citep{BertrandeLis-15, DelgadoMena-17a, Suarez-andres-17}. To obtain the 
'empirical' abundances of other stars we proceed as follows. We first searched for stellar analogs\footnote{We defined stellar analogs as stars with [Fe/H]$\pm$0.1 dex,
\teff$\pm$500K, and $\log$g$\pm$0.3 dex.} for each star in these catalogs: \citet[][]{Suarez-andres-17} for Carbon abundance and \citet{DelgadoMena-17a} for abundances of Mg and Si.
The mean abundance of all the analogs was used as a proxy for the 'empirical' abundance for a given star, and the standard deviation (star-to-star scatter) of the abundances was used as an
error of the empirical' abundance. Oxygen abundance was derived from the empirical formula between [O/H] and [Fe/H] provided in \citet[][]{Suarez-andres-17} which is based on the 
\citet{BertrandeLis-15} data. Original and 'empirical' abundances of the stars are presented in the  Table\,\ref{tab:table1}. As can be seen from the table, the difference between
'empirical' and original abundances can be as large as 0.1 dex. Thus we stress that the 'empirical' abundances should be considered only as rough estimates.

\section{Composition of the planet building blocks}

The model presented in \citet{Santos-15} uses atomic abundances of O, C, Mg, Si and Fe, as input, and with simple stoichometric equations calculates the mass fraction 
of H$_{2}$O, CH$_{4}$, Fe, MgSiO$_{3}$, Mg$_{2}$SiO$_{4}$, the total mass percentage of all heavy elements (Z), the iron mass 
fraction ($f_{iron}$ = m$_{Fe}$/(m$_{Fe}$ + m$_{MgSiO_{3}}$ + m$_{Mg_{2}SiO_{4}}$) and the water mass fraction ($w_{f}$ = m$_{H_{2}O}$/(m$_{H_{2}O}$ + m$_{Fe}$ + m$_{MgSiO_{3}}$ + m$_{Mg_{2}SiO_{4}}$)). 
These values are derived for each star using the original spectroscopic and 'empirical' abundances.
The results are presented in Table\,\ref{tab:table2}. From the table we can see that for the three stars for which together with the 'empirical' abundances 
spectroscopic abundances are available (HD40307, HD10700, and Kepler-22), the derived values are similar and agree within the error bars. However, it should be mention the the uncertainties
of some of the parameters are large, especially if they are derived from the 'empirical' abundances.

\section{Results and Discussion}

Our results summarized in Table\,\ref{tab:table2} show that if small-size and low-mass planets are found in the HZ of the studied seven stars then they are expected to have significantly different 
iron-to-silicate and water mass fractions. In particular, the iron mass fraction in five out of seven cases is significantly lower (from $\sim$24 to $\sim$28\%) than what this model would 
predict for solar-system planet building blocks
i.e. $f_{iron}$ = 33\% Santos et al. (2017, submitted). Water content would also vary from system to system between $\sim$56 to 72\%. Here we should stress again the large 
uncertainties for this parameter that mostly come from the larger errors on the C and O abundances. Note that the model predicts a $w_{f}$ = 60\% which is compatible
with the value of $\sim$67\% derived in \citet{Lodders-03}.

Very recently, Santos et al. (2017, submitted) compiled chemical abundances for large sample of solar-type stars from the solar vicinity and derived the expected composition of the 
planet building blocks. The authors found that stars belonging to different galactic stellar populations (thin disk, thick disk, halo, and high-$\alpha$ metal-rich - \citet{Adibekyan-11})
are expected to have rocky planets with significantly different iron mass and water mass fractions. Our results go well in line with the findings of Santos et al. (2017, submitted), 
since stars in our small sample having different metallicities and ages probably belong to different galactic populations. The results also somehow confirm the prediction of \citet{Adibekyan-16}
that exoplanets in the HZ may have composition different from that of our Earth.

\section{A laconic conclusion}

We estimated the water and iron-to-silicate mass fraction of planet building blocks for seven solar-type stars with precise spectroscopic metallicities that are known to have planets in the HZ.
Our, very simplified analysis show that if rocky planets are found orbiting around these stars they might have different composition when compared with our own planet. To confidently 
answer to the question postulated in the title of this manuscript a far more sophisticated analysis  for each individual object is needed with an important requirement of having 
very precise masses and radius of the planets and very accurate  chemical abundances of the host star \citep[e.g.][]{Dorn-17}.

\section*{Acknowledgments}

{V.A. thanks the organizers of EWASS Special Session 4 (2017), Emeline Bolmont \& Sergi Blanco-Cuaresma, for a very interesting session and for selecting his oral contribution. 
This work was supported by Funda\c{c}\~ao para a Ci\^encia e Tecnologia (FCT) through national funds (ref. PTDC/FIS-AST/7073/2014
and ref. PTDC/FIS-AST/1526/2014) through national funds and by FEDER through COMPETE2020 (ref. POCI-01-0145-FEDER-016880 and ref. POCI-01-0145-FEDER-016886).
V.A., P.F.,  and N.C.S. also acknowledge the support from FCT through Investigador FCT contracts of reference \seqsplit{IF/00650/2015/CP1273/CT0001,} 
\seqsplit{IF/01037/2013/CP1191/CT0001,} and \seqsplit{IF/00169/2012/CP0150/CT0002,} respectively, and POPH/FSE (EC) by FEDER funding through 
the program ``Programa Operacional de Factores de Competitividade - COMPETE''. 
PF further acknowledges support from FCT in the form of an exploratory project of reference IF/01037/2013CP1191/CT0001.}

\bibliographystyle{ewass_ss4proc}
\bibliography{references.bib}

\begin{thebibliography}{46}
\providecommand{\natexlab}[1]{#1}

\bibitem[\protect\astroncite{{Adibekyan}}{2017}]{Adibekyan-17b}
{Adibekyan}, V. 2017, In \emph{Astronomical Society of the Pacific Conference
  Series}, edited by A.~M. {Mickaelian}, H.~A. {Harutyunian}, \& E.~H.
  {Nikoghosyan}, \emph{Astronomical Society of the Pacific Conference Series},
  vol. 511, p.~70.

\bibitem[\protect\astroncite{{Adibekyan} \emph{et~al.}}{2016}]{Adibekyan-16}
{Adibekyan}, V., {Figueira}, P., \& {Santos}, N.~C. 2016, Origins of Life and
  Evolution of the Biosphere, 46, 351.

\bibitem[\protect\astroncite{{Adibekyan} \emph{et~al.}}{2017}]{Adibekyan-17}
{Adibekyan}, V., {Gon{\c c}alves da Silva}, H.~M., {Sousa}, S.~G., {Santos},
  N.~C., {Delgado Mena}, E., \emph{et~al.} 2017, Astrophysics, 60, 325.

\bibitem[\protect\astroncite{{Adibekyan} \emph{et~al.}}{2015}]{Adibekyan-15}
{Adibekyan}, V., {Santos}, N.~C., {Figueira}, P., {Dorn}, C., {Sousa}, S.~G.,
  \emph{et~al.} 2015, \aap, 581, L2.

\bibitem[\protect\astroncite{{Adibekyan}
  \emph{et~al.}}{2012{\natexlab{a}}}]{Adibekyan-12a}
{Adibekyan}, V.~Z., {Delgado Mena}, E., {Sousa}, S.~G., {Santos}, N.~C.,
  {Israelian}, G., \emph{et~al.} 2012{\natexlab{a}}, \aap, 547, A36.

\bibitem[\protect\astroncite{{Adibekyan}
  \emph{et~al.}}{2013{\natexlab{a}}}]{Adibekyan-13b}
{Adibekyan}, V.~Z., {Figueira}, P., {Santos}, N.~C., {Hakobyan}, A.~A.,
  {Sousa}, S.~G., \emph{et~al.} 2013{\natexlab{a}}, \aap, 554, A44.

\bibitem[\protect\astroncite{{Adibekyan}
  \emph{et~al.}}{2013{\natexlab{b}}}]{Adibekyan-13}
{Adibekyan}, V.~Z., {Figueira}, P., {Santos}, N.~C., {Mortier}, A.,
  {Mordasini}, C., \emph{et~al.} 2013{\natexlab{b}}, \aap, 560, A51.

\bibitem[\protect\astroncite{{Adibekyan} \emph{et~al.}}{2011}]{Adibekyan-11}
{Adibekyan}, V.~Z., {Santos}, N.~C., {Sousa}, S.~G., \& {Israelian}, G. 2011,
  \aap, 535, L11.

\bibitem[\protect\astroncite{{Adibekyan}
  \emph{et~al.}}{2012{\natexlab{b}}}]{Adibekyan-12b}
{Adibekyan}, V.~Z., {Santos}, N.~C., {Sousa}, S.~G., {Israelian}, G., {Delgado
  Mena}, E., \emph{et~al.} 2012{\natexlab{b}}, \aap, 543, A89.

\bibitem[\protect\astroncite{{Adibekyan}
  \emph{et~al.}}{2012{\natexlab{c}}}]{Adibekyan-12c}
{Adibekyan}, V.~Z., {Sousa}, S.~G., {Santos}, N.~C., {Delgado Mena}, E.,
  {Gonz{\'a}lez Hern{\'a}ndez}, J.~I., \emph{et~al.} 2012{\natexlab{c}}, \aap,
  545, A32.

\bibitem[\protect\astroncite{{Beaug{\'e}} \& {Nesvorn{\'y}}}{2013}]{Beauge-13}
{Beaug{\'e}}, C. \& {Nesvorn{\'y}}, D. 2013, \apj, 763, 12.

\bibitem[\protect\astroncite{{Bensby} \emph{et~al.}}{2003}]{Bensby-13}
{Bensby}, T., {Feltzing}, S., \& {Lundstr{\"o}m}, I. 2003, \aap, 410, 527.

\bibitem[\protect\astroncite{{Bertran de Lis}
  \emph{et~al.}}{2015}]{BertrandeLis-15}
{Bertran de Lis}, S., {Delgado Mena}, E., {Adibekyan}, V.~Z., {Santos}, N.~C.,
  \& {Sousa}, S.~G. 2015, \aap, 576, A89.

\bibitem[\protect\astroncite{{Borucki} \emph{et~al.}}{2013}]{Borucki-13}
{Borucki}, W.~J., {Agol}, E., {Fressin}, F., {Kaltenegger}, L., {Rowe}, J.,
  \emph{et~al.} 2013, Science, 340, 587.

\bibitem[\protect\astroncite{{Borucki} \emph{et~al.}}{2012}]{Borucki-12}
{Borucki}, W.~J., {Koch}, D.~G., {Batalha}, N., {Bryson}, S.~T., {Rowe}, J.,
  \emph{et~al.} 2012, \apj, 745, 120.

\bibitem[\protect\astroncite{{Buchhave} \& {Latham}}{2015}]{Buchhave-15}
{Buchhave}, L.~A. \& {Latham}, D.~W. 2015, \apj, 808, 187.

\bibitem[\protect\astroncite{{Campante} \emph{et~al.}}{2015}]{Campante-15}
{Campante}, T.~L., {Barclay}, T., {Swift}, J.~J., {Huber}, D., {Adibekyan},
  V.~Z., \emph{et~al.} 2015, \apj, 799, 170.

\bibitem[\protect\astroncite{{Delgado Mena}
  \emph{et~al.}}{2017{\natexlab{a}}}]{DelgadoMena-17}
{Delgado Mena}, E., {Tsantaki}, M., {Adibekyan}, V.~Z., {Sousa}, S.~G.,
  {Santos}, N.~C., \emph{et~al.} 2017{\natexlab{a}}, [arXiv:1707.05156].

\bibitem[\protect\astroncite{{Delgado Mena}
  \emph{et~al.}}{2017{\natexlab{b}}}]{DelgadoMena-17a}
{Delgado Mena}, E., {Tsantaki}, M., {Adibekyan}, V.~Z., {Sousa}, S.~G.,
  {Santos}, N.~C., \emph{et~al.} 2017{\natexlab{b}}, [arXiv:1705.04349].

\bibitem[\protect\astroncite{{Dorn} \emph{et~al.}}{2017}]{Dorn-17}
{Dorn}, C., {Hinkel}, N.~R., \& {Venturini}, J. 2017, \aap, 597, A38.

\bibitem[\protect\astroncite{{Dorn} \emph{et~al.}}{2015}]{Dorn-15}
{Dorn}, C., {Khan}, A., {Heng}, K., {Connolly}, J.~A.~D., {Alibert}, Y.,
  \emph{et~al.} 2015, \aap, 577, A83.

\bibitem[\protect\astroncite{{Gonzalez}}{1997}]{Gonzalez-97}
{Gonzalez}, G. 1997, Mon Not R Astron Soc, 285, 403.

\bibitem[\protect\astroncite{{Haywood}}{2009}]{Haywood-09}
{Haywood}, M. 2009, \apjl, 698, L1.

\bibitem[\protect\astroncite{{Haywood} \emph{et~al.}}{2013}]{Haywood-13}
{Haywood}, M., {Di Matteo}, P., {Lehnert}, M.~D., {Katz}, D., \& {G{\'o}mez},
  A. 2013, \aap, 560, A109.

\bibitem[\protect\astroncite{{Jenkins} \emph{et~al.}}{2015}]{Jenkins-15}
{Jenkins}, J.~M., {Twicken}, J.~D., {Batalha}, N.~M., {Caldwell}, D.~A.,
  {Cochran}, W.~D., \emph{et~al.} 2015, \aj, 150, 56.

\bibitem[\protect\astroncite{{Johnson} \& {Li}}{2012}]{Johnson-12}
{Johnson}, J.~L. \& {Li}, H. 2012, \apj, 751, 81.

\bibitem[\protect\astroncite{{Kordopatis} \emph{et~al.}}{2015}]{Kordopatis-15}
{Kordopatis}, G., {Wyse}, R.~F.~G., {Gilmore}, G., {Recio-Blanco}, A., {de
  Laverny}, P., \emph{et~al.} 2015, \aap, 582, A122.

\bibitem[\protect\astroncite{{Lodders}}{2003}]{Lodders-03}
{Lodders}, K. 2003, \apj, 591, 1220.

\bibitem[\protect\astroncite{{Mortier} \emph{et~al.}}{2013}]{Mortier-13}
{Mortier}, A., {Santos}, N.~C., {Sousa}, S., {Israelian}, G., {Mayor}, M.,
  \emph{et~al.} 2013, \aap, 551, A112.

\bibitem[\protect\astroncite{{Morton} \emph{et~al.}}{2016}]{Morton-16}
{Morton}, T.~D., {Bryson}, S.~T., {Coughlin}, J.~L., {Rowe}, J.~F.,
  {Ravichandran}, G., \emph{et~al.} 2016, \apj, 822, 86.

\bibitem[\protect\astroncite{{Mulders} \emph{et~al.}}{2016}]{Mulders-16}
{Mulders}, G.~D., {Pascucci}, I., {Apai}, D., {Frasca}, A., \&
  {Molenda-{\.Z}akowicz}, J. 2016, \aj, 152, 187.

\bibitem[\protect\astroncite{{Nissen} \emph{et~al.}}{2017}]{Nissen-17}
{Nissen}, P.~E., {Silva Aguirre}, V., {Christensen-Dalsgaard}, J., {Collet},
  R., {Grundahl}, F., \emph{et~al.} 2017, ArXiv e-prints.

\bibitem[\protect\astroncite{{Recio-Blanco}
  \emph{et~al.}}{2014}]{Recio-Blanco-14}
{Recio-Blanco}, A., {de Laverny}, P., {Kordopatis}, G., {Helmi}, A., {Hill},
  V., \emph{et~al.} 2014, \aap, 567, A5.

\bibitem[\protect\astroncite{{Rowe} \emph{et~al.}}{2014}]{Rowe-14}
{Rowe}, J.~F., {Bryson}, S.~T., {Marcy}, G.~W., {Lissauer}, J.~J.,
  {Jontof-Hutter}, D., \emph{et~al.} 2014, \apj, 784, 45.

\bibitem[\protect\astroncite{{Santos} \emph{et~al.}}{2017}]{Santos-17}
{Santos}, N.~C., {Adibekyan}, V., {Figueira}, P., {Andreasen}, D.~T., {Barros},
  S.~C.~C., \emph{et~al.} 2017, \aap, 603, A30.

\bibitem[\protect\astroncite{{Santos} \emph{et~al.}}{2015}]{Santos-15}
{Santos}, N.~C., {Adibekyan}, V., {Mordasini}, C., {Benz}, W., {Delgado-Mena},
  E., \emph{et~al.} 2015, \aap, 580, L13.

\bibitem[\protect\astroncite{{Santos} \emph{et~al.}}{2001}]{Santos-01}
{Santos}, N.~C., {Israelian}, G., \& {Mayor}, M. 2001, Astron Astrophys, 373,
  1019.

\bibitem[\protect\astroncite{{Schuler} \emph{et~al.}}{2015}]{Schuler-15}
{Schuler}, S.~C., {Vaz}, Z.~A., {Katime Santrich}, O.~J., {Cunha}, K., {Smith},
  V.~V., \emph{et~al.} 2015, \apj, 815, 5.

\bibitem[\protect\astroncite{{Sousa} \emph{et~al.}}{2011}]{Sousa-11}
{Sousa}, S.~G., {Santos}, N.~C., {Israelian}, G., {Mayor}, M., \& {Udry}, S.
  2011, Astron Astrophys, 533, A141.

\bibitem[\protect\astroncite{{Sousa} \emph{et~al.}}{2008}]{Sousa-08}
{Sousa}, S.~G., {Santos}, N.~C., {Mayor}, M., {Udry}, S., {Casagrande}, L.,
  \emph{et~al.} 2008, \aap, 487, 373.

\bibitem[\protect\astroncite{{Su{\'a}rez-Andr{\'e}s}
  \emph{et~al.}}{2017}]{Suarez-andres-17}
{Su{\'a}rez-Andr{\'e}s}, L., {Israelian}, G., {Gonz{\'a}lez Hern{\'a}ndez},
  J.~I., {Adibekyan}, V.~Z., {Delgado Mena}, E., \emph{et~al.} 2017, \aap, 599,
  A96.

\bibitem[\protect\astroncite{{Thiabaud} \emph{et~al.}}{2014}]{Thiabaud-14}
{Thiabaud}, A., {Marboeuf}, U., {Alibert}, Y., {Cabral}, N., {Leya}, I.,
  \emph{et~al.} 2014, Astron Astrophys, 562, A27.

\bibitem[\protect\astroncite{{Torres} \emph{et~al.}}{2015}]{Torres-15}
{Torres}, G., {Kipping}, D.~M., {Fressin}, F., {Caldwell}, D.~A., {Twicken},
  J.~D., \emph{et~al.} 2015, \apj, 800, 99.

\bibitem[\protect\astroncite{{Tsantaki} \emph{et~al.}}{2013}]{Tsantaki-13}
{Tsantaki}, M., {Sousa}, S.~G., {Adibekyan}, V.~Z., {Santos}, N.~C., {Mortier},
  A., \emph{et~al.} 2013, \aap, 555, A150.

\bibitem[\protect\astroncite{{Turbo-King} \emph{et~al.}}{2017}]{Turbo-King-17}
{Turbo-King}, M., {Tang}, B.~R., {Habeertable}, Z., {Chouffe}, M.~C.,
  {Exquisit}, B., \emph{et~al.} 2017, [arXiv:1703.10803].

\bibitem[\protect\astroncite{{Zhu} \emph{et~al.}}{2016}]{Zhu-16}
{Zhu}, W., {Wang}, J., \& {Huang}, C. 2016, \apj, 832, 196.

\end{thebibliography}

\end{document}